\documentclass[%
reprint,
superscriptaddress,
 amsmath,amssymb,
 aps,
floatfix,
]{revtex4-2}

\usepackage{graphicx}
\usepackage{dcolumn}
\usepackage{bm}
\usepackage{placeins}
\usepackage{fancyvrb}
\usepackage[caption=false]{subfig}
\usepackage[version=3]{mhchem}

\begin{document}

\title{High-Pressure Structural Evolution of Disordered Polymeric CS\textsubscript{2}}

\author{Jinwey Yan}
\affiliation{Key Laboratory of Materials Physics, Institute of Solid State Physics, Chinese Academy of Sciences, Hefei, 230031, China}
\affiliation{University of Science and Technology of China, Hefei 230026, China}

\author{Ondrej T\'{o}th}
\affiliation{Department of Experimental Physics, Comenius University, 
Mlynsk\'{a} Dolina F2, 842 48 Bratislava, Slovakia}

\author{Wan Xu}
\affiliation{Key Laboratory of Materials Physics, Institute of Solid State Physics, Chinese Academy of Sciences, Hefei, 230031, China}
\affiliation{University of Science and Technology of China, Hefei 230026, China}

\author{Xiao-Di Liu}
\affiliation{Key Laboratory of Materials Physics, Institute of Solid State Physics, Chinese Academy of Sciences, Hefei, 230031, China}

\author{Eugene Gregoryanz}
\affiliation{Key Laboratory of Materials Physics, Institute of Solid State Physics, Chinese Academy of Sciences, Hefei, 230031, China}
\affiliation{School of Physics and Astronomy and Centre for Science at Extreme Conditions, University of Edinburgh, Edinburgh EH9 3JZ, UK}
\affiliation{Center for High Pressure Science Technology Advanced Research, 1690 Cailun Road, Shanghai, 201203, China}

\author{Philip Dalladay-Simpson}
\affiliation{Center for High Pressure Science Technology Advanced Research, 1690 Cailun Road, Shanghai, 201203, China}

\author{Zeming Qi}
\affiliation{National Synchrotron Radiation Laboratory, University of Science and Technology of China, Hefei 230029, China}

\author{Shiyu Xie}
\affiliation{National Synchrotron Radiation Laboratory, University of Science and Technology of China, Hefei 230029, China}

\author{Federico Gorelli}
\affiliation{Key Laboratory of Materials Physics, Institute of Solid State Physics, Chinese Academy of Sciences, Hefei, 230031, China}
\affiliation{Center for High Pressure Science Technology Advanced Research, 1690 Cailun Road, Shanghai, 201203, China}
\affiliation{Istituto Nazionale di Ottica (CNR-INO) and European Laboratory for non Linear Spectroscopy (LENS), via N. Carrara 1, 50019 Sesto Fiorentino, Italy}

\author{Roman Marto\v{n}\'{a}k}
\affiliation{Department of Experimental Physics, Comenius University, 
Mlynsk\'{a} Dolina F2, 842 48 Bratislava, Slovakia}

\author{Mario Santoro}
\affiliation{Key Laboratory of Materials Physics, Institute of Solid State Physics, Chinese Academy of Sciences, Hefei, 230031, China}
\affiliation{Istituto Nazionale di Ottica (CNR-INO) and European Laboratory for non Linear Spectroscopy (LENS), via N. Carrara 1, 50019 Sesto Fiorentino, Italy}

\date{\today}

\begin{abstract}
Carbon disulfide, \ce{CS2}, is an archetypal double-bonded molecular system belonging to the rich class of group IV-group VI, \ce{AB2} compounds. It is widely and since long time believed that upon compression at several GPa a polymeric chain of type (-(C=S)-S-)\textsubscript{n} named Bridgman's black polymer will form. By combining optical spectroscopy and synchrotron X-ray diffraction data with ab initio simulations, we demonstrate that the structure of the Bridgman’s black polymer is remarkably different. Solid molecular \ce{CS2} undergoes a pressure-induced structural transformation at around 10-11~GPa, developing a disordered polymeric system. The polymer consists of 3-fold and 4-fold coordinated carbon atoms with an average carbon coordination continuously increasing upon further compression to 40~GPa. Polymerization also gives rise to some C=C double bonds. Upon decompression, the structural changes are partially reverted, a very small amount of molecular \ce{CS2} is recovered, while the sample undergoes partial chemical disproportionation. Our work uncovers the non-trivial high-pressure structural evolution in one of the simplest molecular systems exhibiting molecular as well as polymeric phases.
\end{abstract}

\maketitle

\section*{Introduction}
Carbon disulfide belongs to the important class of \mbox{IV-VI}, \ce{AB2} compounds. The general high pressure trend of these compounds is to increase the coordination of the cation by the anion in the GPa/tens of GPa pressure range. Indeed, it is well established that the coordination goes from 4 to 6 in \ce{SiO2} \cite{SiO2, Prokopenko, HainesQuartz}, \ce{GeO2} \cite{Durben, Guthrie} and \ce{SiS2} \cite{PlasienkaSiS2, Evers, Wang}, and from 2 to mixed 3-4 and pure 4 in non-molecular amorphous and crystalline \ce{CO2}, respectively \cite{Iota, SantoroA, Montoya, SantoroB, Datchi, PlasienkaCO2, Shieba} while, at hundreds of GPa, \ce{CO2} is also predicted to exhibit 6-fold coordination \cite{CO2_sixfold}. At variance with the above mentioned systems, \ce{CS2} is a strongly metastable substance, even in the molecular state at ambient conditions. Several works reported the high-pressure formation of extended solids obtained by compressing molecular \ce{CS2} above several GPa \cite{Bridgman, Agnew, Dias}, yet the chemical nature and structure of these materials is still very elusive, which implies an important knowledge gap in the class of IV-VI, \ce{AB2} compounds. IR spectroscopy investigations showed that \ce{CS2} chemically transforms above 8.3~GPa, at room temperature, and the product was indicated being a mixture of the so-called Bridgman’s 1D black polymer, (-(C=S)-S-)\textsubscript{n}, where C is in planar 3-fold coordination by S, and \ce{CS2} dimers \cite{Agnew}. This interpretation was not grounded on quantitative structural models and it was based on a limited set of data. More recently, \ce{CS2} was investigated up to 90~GPa by Raman spectroscopy, X-ray diffraction (XRD) and ab initio calculations \cite{Dias}, confirming the polymerization at around 9~GPa and arguing on an additional structural transformation at about 30~GPa. These two forms were shown to be opaque and disordered and claimed to have C in 3-fold and 4-fold coordination, respectively. On the other hand, Raman spectra hardly demonstrated any major changes above 9~GPa, and disordered structural models were still lacking. Subsequently, a constrained first-principles evolutionary search was conducted up to 200~GPa in order to identify the lowest-enthalpy structures of non-molecular \ce{CS2} with C in 4-fold coordination by S \cite{Naghavi2015}, yet the structure of the experimentally obtained extended, disordered \ce{CS2} remained unsolved. In another ab initio simulation study with constrained C:S=1:2 stoichiometry, the molecular crystal  has been predicted to transform into non-molecular either amorphous or crystalline solids, above 10~GPa \cite{Zarifi}. All these materials include C-C and S-S bonds along with C-S bonds, and also C and S are predicted to separate on the spatial scale of the simulation cell. In this work too, the comparison to experimental investigations is limited and indirect. It should be also noted that the application of crystal structure prediction (CSP) techniques to a system prone to decomposition is in principle non-trivial, since a whole spectrum of structures can be found, depending on the degree of separation of the two elements (unless constraints are imposed).

In order to provide a deeper insight into disordered polymeric \ce{CS2} and unveil the nature of this elusive material, we conducted a combined experimental and computational study, based on Raman and IR spectroscopy, synchrotron X-ray diffraction (XRD) and ab initio molecular dynamics simulations and crystal structure prediction. In the following, we will present our investigation of polymeric \ce{CS2} at 0-40~GPa. We will show that this material is far more complex than the simple Bridgman’s polymer, since it consists of a strongly pressure-dependent mixture of C sites in 3-fold and 4-fold coordination by S as well as chains with C=C double bonds.

\section*{Results and Discussion}
\begin{figure}[hbt!]
\centering
  \includegraphics[height=15.4cm]{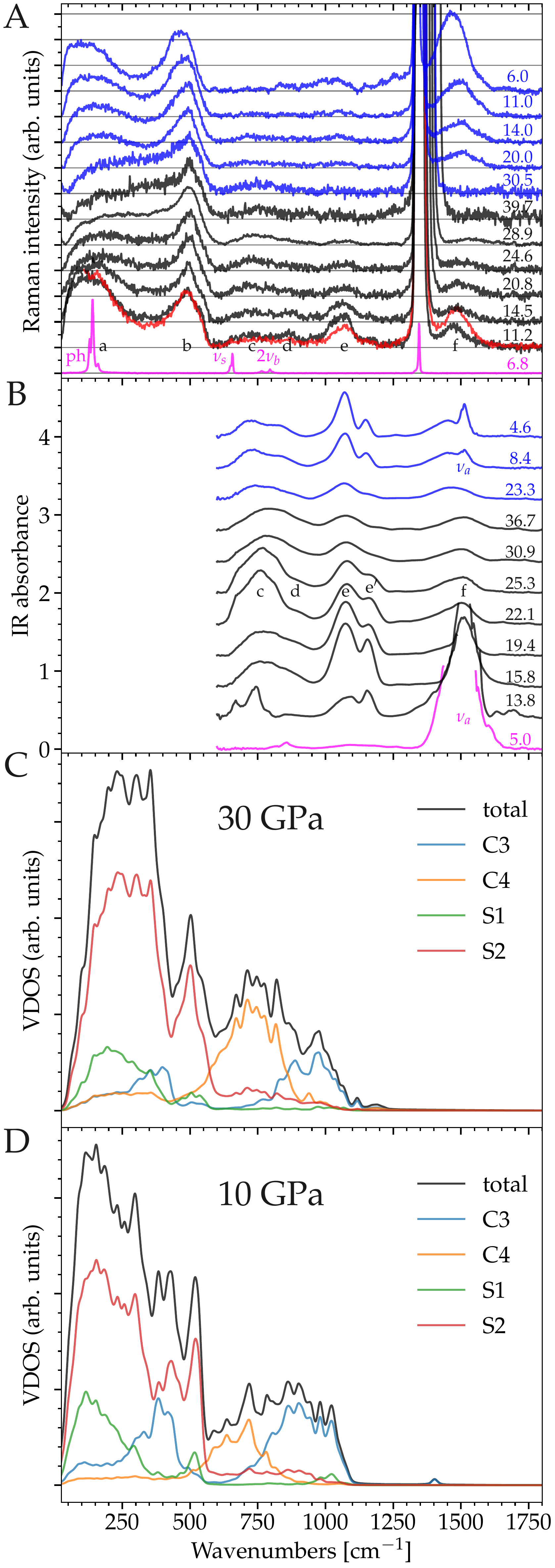}
  \caption{Panel A and B, magenta, black and blue lines: selected experimental Raman A and IR B spectra of \ce{CS2}, measured upon increasing (magenta and black) and subsequently decreasing (blue) pressure. Laser wavelength for Raman: 660~nm. Sharp ph, $\nu_s$ and 2$\nu_b$ peaks in Raman and strong saturating $\nu_a$ peak in IR: lattice, symmetric molecular stretching, double (overtone) molecular bending, and asymmetric molecular stretching modes of the \ce{CS2} molecular crystal, respectively. Broad \textit{a}-\textit{f} peaks: peaks of polymeric \ce{CS2}. Red line: Raman spectrum measured on a similar sample using the 532~nm laser line. For the sake of clarity, Raman spectra of polymeric \ce{CS2} have been vertically rescaled by the same intensity of peak \textit{b}, while the Raman spectrum of molecular \ce{CS2} is rescaled by a factor of about 10. Raman and IR spectra have also been vertically shifted. Panel C and D: total and partial vibrational density of states (VDOS) calculated for two different pressures from configurations obtained in ab initio MD. Lines of different colors indicate different C and S sites (see text).}
  \label{fig:1}
\end{figure}

Optical, vibrational spectroscopies are the key tool for a clear identification of chemical species and pressure-driven changes in the chemical nature of the sample. In figure~\ref{fig:1}A, we report selected Raman spectra of solid \ce{CS2} measured upon increasing and decreasing pressure in the 6-40~GPa pressure range. Since \ce{CS2} is photosensitive, particularly in the molecular phase, we have strongly limited the laser power to 1~mW or less for a beam spot of about 2~$\mu$m and the acquisition time was also limited to 1~second in the molecular phase, and tens of seconds in the polymeric form. In the GPa range, we only observe sharp and intense peaks related to the molecular and lattice modes of the \ce{CS2} crystal. Above 9-10~GPa, the sample becomes entirely opaque, black, and the spectrum modifies significantly, abruptly and irreversibly: the sharp peaks are entirely replaced by at least six much broader and weaker bands marked as: \textit{a}, \textit{b}, \textit{c}, \textit{d}, \textit{e}, and \textit{f} in figure~\ref{fig:1}A, at about 130~cm\textsuperscript{-1}, 490~cm\textsuperscript{-1}, 740-750~cm\textsuperscript{-1}, 870~cm\textsuperscript{-1}, 1060~cm\textsuperscript{-1} and 1480~cm\textsuperscript{-1}, respectively, signaling a major chemical modification such as the formation of a polymeric and likely disordered form. We note that previous Raman investigations reported only band \textit{b} \cite{Dias}. In our case, the very same spectrum is observed by using two distinct laser wavelengths, 660~nm and 532~nm, which confirms that \textit{a}-\textit{f} bands are all true Raman peaks, i.e. none of them is a fluorescence peak. These new bands undergo partially reversible intensity changes along the pressure cycle. Indeed, band \textit{a} decreases substantially upon increasing pressure to 40~GPa, and it increases back when pressure is reduced. Similarly, bands \textit{e} and \textit{f} reversibly decrease upon increasing pressure, and band \textit{f} becomes the dominant peak upon returning to 6~GPa. Bands \textit{b}, \textit{c} and \textit{d} are always observed along the pressure cycle, except upon
decreasing pressure below 14~GPa where \textit{c} and \textit{d} become hardly detectable. In figure~\ref{fig:1}A, we report selected medium IR absorption spectra of \ce{CS2} measured at 600-1800~cm\textsuperscript{-1} along a typical pressure cycle in the 4-37~GPa range. Comparison to Raman spectra clearly shows that the two types of spectra for polymeric \ce{CS2}, IR and Raman, have very similar bands, in the common frequency range, while the IR bands are observed with a much higher signal to noise ratio. We then adopt the same Raman labels for the main IR peaks of polymeric \ce{CS2}: \textit{c}, \textit{d}, and \textit{f}. Similarly to Raman, the IR spectrum has entirely molecular origin at 5~GPa, as testified by the strong saturating peak at around 1500~cm\textsuperscript{-1} assigned to the antisymmetric \ce{CS2} stretching mode. The spectrum undergoes sudden and major changes upon increasing pressure above 9-10~GPa, pointing to the formation of a polymeric product. In fact, the molecular peak is replaced by non-molecular broad bands \textit{c}, \textit{e}, \textit{e’} (\textit{e’} not seen in the Raman spectrum) and \textit{f}. At a closer glance, band \textit{c} exhibits multiple structure at some pressures, while bands \textit{e} and \textit{e’} reversibly decrease upon increasing pressure with \textit{e’} almost entirely vanishing at the highest pressures. Remnants of the molecular peak persist up to about 20~GPa, partially overlapping to the polymeric \textit{f} band. Then, the molecular peak forms back very slightly by reducing pressure below 10~GPa, indicating the backformation of a very small amount, as compared to the initial sample, of molecular \ce{CS2} not seen through the much noisier Raman spectra. Interestingly, bands \textit{e} and \textit{f} already provide a few heuristic hints on the chemical nature of polymeric \ce{CS2}. In fact, band \textit{e} is compatible with the IR peak observed previously (\cite{Agnew} and references therein) and empirically assigned to the C=S stretching mode of the Bridgman’s polymer, \mbox{(-(C=S)-S-)\textsubscript{n}}, while band \textit{f} indicates the presence of C=C double bonds. Therefore, bands \textit{e} and \textit{f} point to C in planar, 3-fold coordination, which appears suddenly above the polymerization pressure and then tends to reversibly decrease upon increasing pressure.

\begin{figure}
 \centering
 \includegraphics[height=10.5cm]{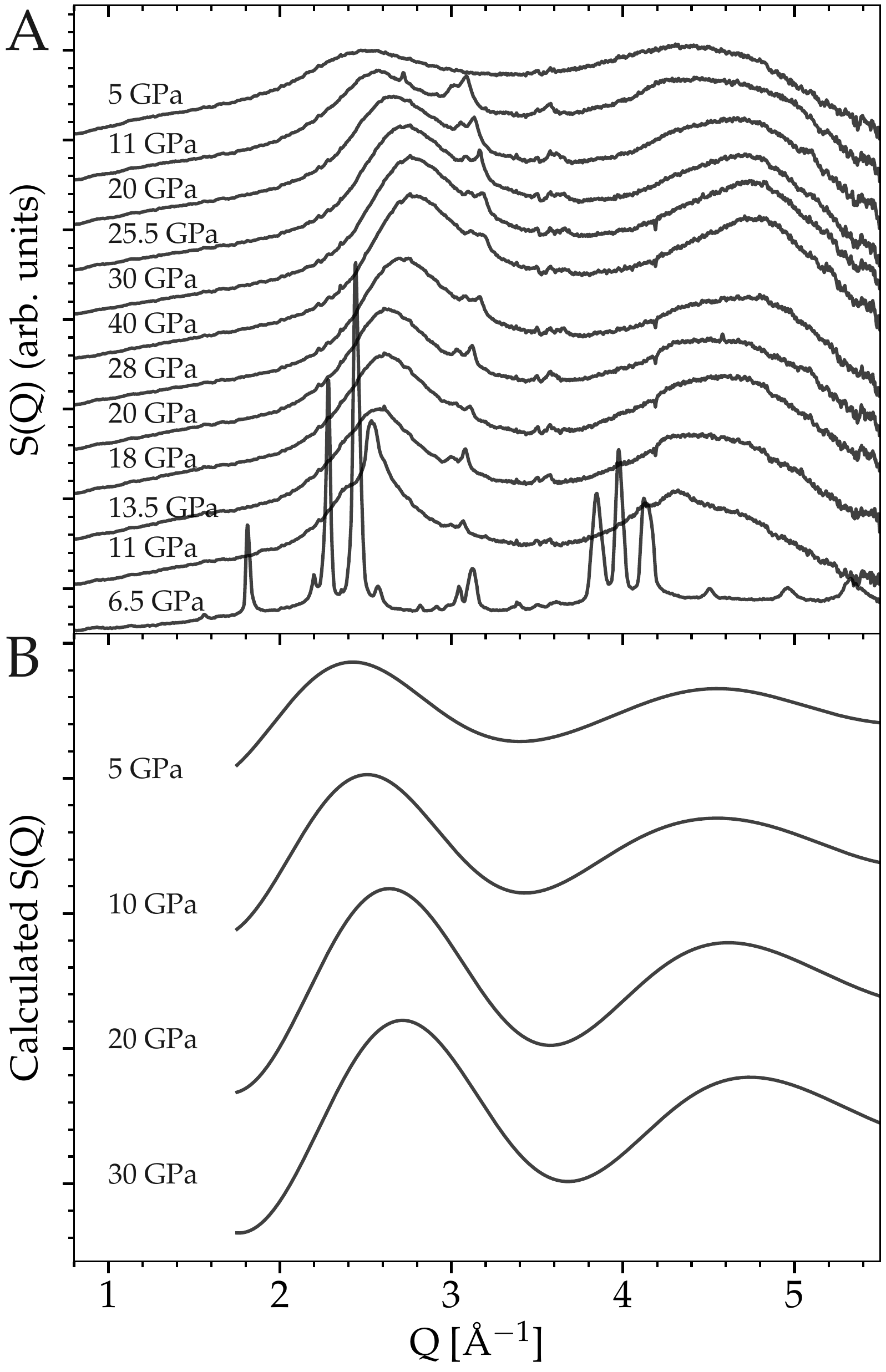}
 \includegraphics[height=10.5cm]{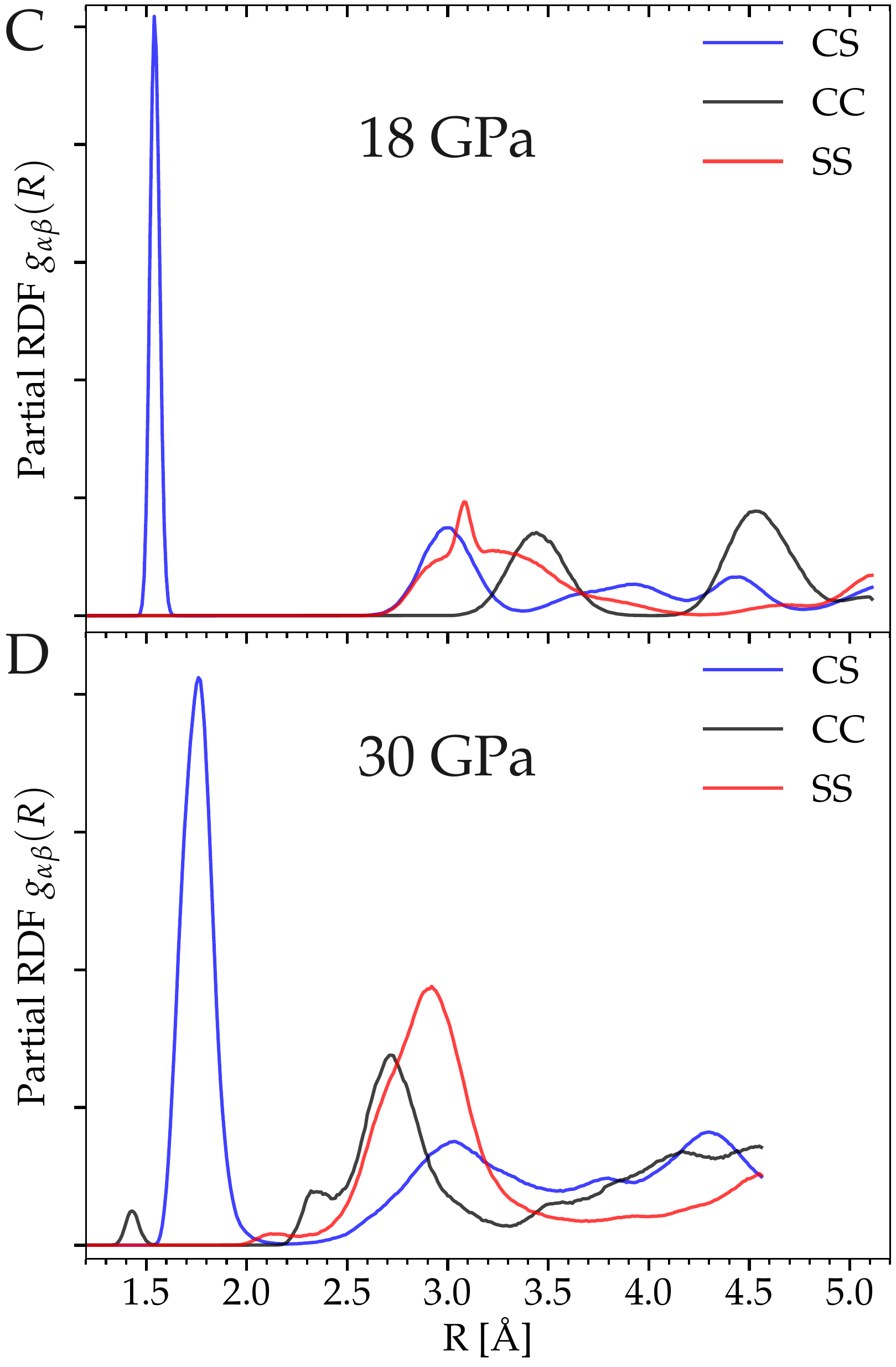}
 \caption{Panel A: selected S(Q) patterns of solid \ce{CS2} measured, through XRD, along a typical pressure cycle: 6.5-40.5~GPa. Panel B: S(Q) obtained from ab initio MD simulations. Panel C and D: computationally (MD) obtained partial Radial Distribution Functions, RDFs, calculated for the \textit{Cmce} molecular crystal at 18~GPa C and for the polymeric sample at 30~GPa D.}
 \label{fig:2}
\end{figure}

For directly proving the disordered character of polymeric \ce{CS2}, we performed synchrotron XRD measurements. In figure~\ref{fig:2}A, we report selected patterns of the static structure factor, S(Q), of solid \ce{CS2} measured along a typical pressure cycle in the 5-40~GPa range. At 6.5~GPa we observe only the sharp Bragg peaks of the \ce{CS2} molecular crystal (\textit{Cmce}). Instead, at 11~GPa, we observe a glassy-like S(Q) dominated by a broad peak at around 2.5~\r{A}\textsuperscript{-1}, that is very close to the still visible strongest Bragg peaks of the molecular phase. The change is complete at 13.5~GPa where, besides the broad diffuse peak, we also observe, at all pressures, weak Bragg peaks at 3.0-3.6~\r{A}\textsuperscript{-1}, whose possible origin will be discussed in the theoretical section. These changes show that \ce{CS2} undergoes a major transformation above 10~GPa, consistently with Raman and IR investigations, and that the novel form is indeed a disordered material. The half width at half maximum of the main diffuse peak is weakly pressure dependent and it amounts to 0.25-0.30~\r{A}\textsuperscript{-1}, which implies a spatial coherence length of about 3.3-4.0~\r{A} or longer. This peak exhibits a normal, reversible shift to higher exchanged momentum Q upon increasing pressure.

\begin{figure}
\centering
  \includegraphics[width=.95\linewidth]{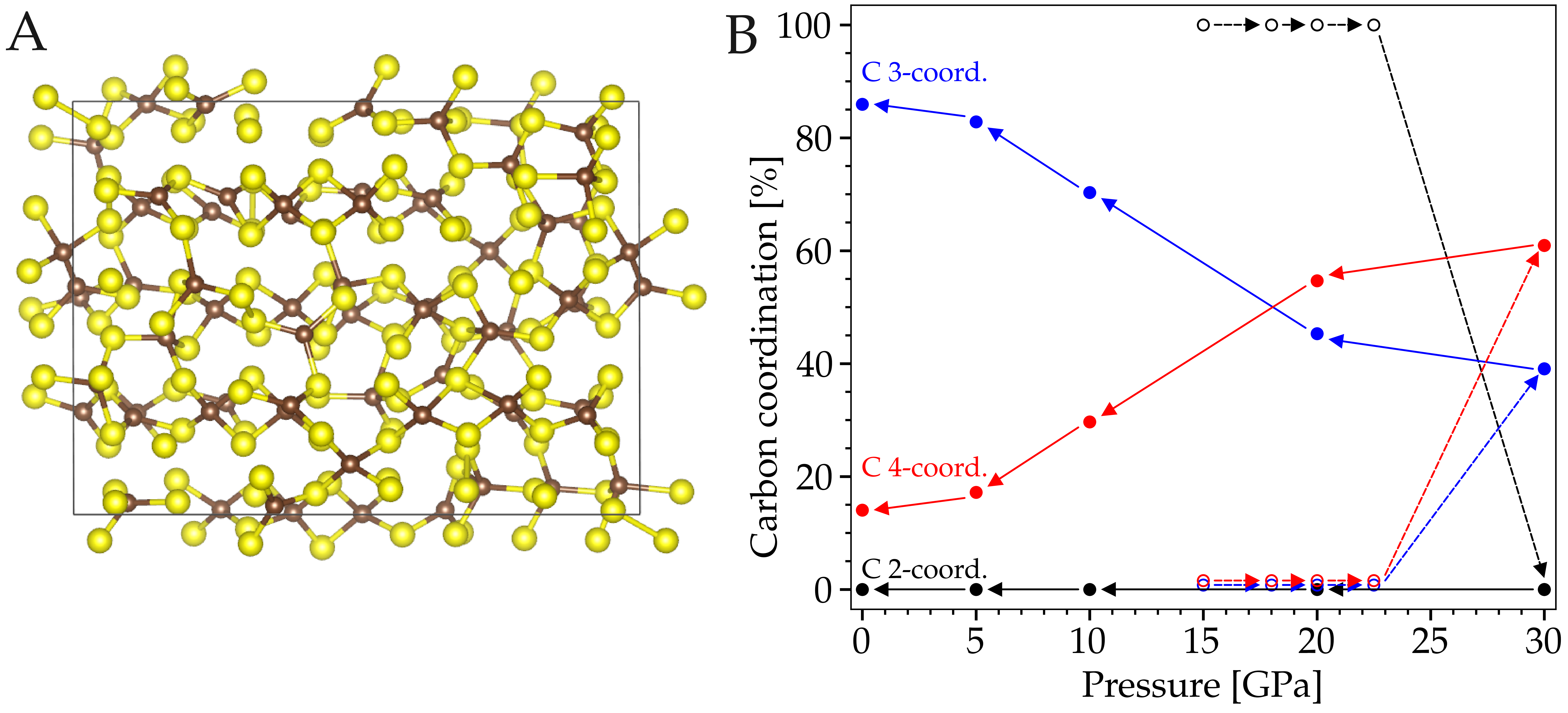}
  \caption{Left: Structure of polymeric \ce{CS2} from ab initio MD after polymerization of the \textit{Cmce} molecular crystal at 30~GPa. Right: fraction of C atoms with different coordination as function of pressure from ab initio MD. Open and full circles are obtained upon increasing and decreasing pressure, respectively.}
  \label{fig:3}
\end{figure}

In order to better understand the structural transformation of the molecular crystal and the subsequent evolution of the disordered polymeric sample upon compression and decompression, we performed ab initio simulations. We started with direct compression of the \textit{Cmce} sample consisting of 192 atoms using ab initio MD. The protocol is shown in figure~S1 (Supp. Mat.). Due to the well-known time-scale-gap problem the sample did not transform at pressure close to the experimental value of 11~GPa and polymerization occurred upon over-pressurization to 30~GPa. The resulting sample is a disordered polymeric system consisting of 3-fold and 4-fold coordinated carbon atoms (figure~\ref{fig:3}A). It can be seen that there is still visible medium-range order in the positions of S atoms, originating from the molecular phase while C atoms seem to be more disordered. This partial order of S atoms might give rise to the sharp Bragg-like peaks observed at 3.1-3.6~\r{A}\textsuperscript{-1} in the experimental S(Q) after polymerization at 11~GPa. The polymerized structure was then decompressed to 20, 10, 5 and 0~GPa in order to check for potential pressure-induced structural changes. Figure~\ref{fig:3}B shows the evolution of the number of C atoms in molecular 2-fold (C2), and in polymeric planar 3-fold (C3) and tetrahedral 4-fold (C4) coordination upon compression and decompression. Right after polymerization at 30~GPa the system contains about 61\% of C4 atoms and 39\% of C3 atoms. Upon decompression, this ratio reverts below 20~GPa and at 5~GPa almost 83\% of C atoms are 3-fold coordinated. In figure~\ref{fig:2}, we show a comparison of the S(Q), calculated from our samples (fig.~\ref{fig:2}B), to the experimental one obtained from XRD (fig.~\ref{fig:2}A). The Q positions of the first two broad peaks of the calculated S(Q) compare fairly well to those of the experimental structure factor and the evolution upon decompression also follows closely the experimental behavior. In figure~\ref{fig:2}, we also report the computationally obtained partial radial distribution functions, RDF, calculated for the \textit{Cmce} molecular crystal at 18~GPa (fig.~\ref{fig:2}C) and for the polymeric sample at 30~GPa (fig.~\ref{fig:2}D). Remarkably, the position of the nearest neighbors’ C-S peak increases from 1.54~\r{A} in the molecular crystal to 1.76~\r{A} in polymeric \ce{CS2}, indicating breaking of the molecular C=S double bonds. Also, the existence of C-C nearest neighbor’s in the polymer, although very limited, is supported by a peak at 1.43~\r{A} in the CC partial RDF.

In order to compare computational results to the experimental Raman and IR spectra we further calculated the vibrational density of states (VDOS) for the disordered samples at 30, 20, 10 and 5~GPa. While a direct comparison of the experimental Raman spectra and VDOS intensities is not possible, VDOS represents a proxy to these spectra, particularly for amorphous materials. The results are shown in figure~\ref{fig:1}C (30~GPa) and \ref{fig:1}D (10~GPa). It is very instructive to analyze the partial contributions to VDOS where one can identify specific signatures of atoms with different coordination and different chemical environment. Based on these graphs we can assign the Raman and IR peaks as follows: peaks \textit{a} and \textit{b} around 130~cm\textsuperscript{-1} and 500~cm\textsuperscript{-1}, respectively, mainly come from S atoms (S2) bridging two C atoms. The broad peak \textit{c} at 600-800~cm\textsuperscript{-1} mainly encodes contributions from 4-fold coordinated C atoms (C4). We note that similarly to the experiment, the intensity of this peak decreases upon decreasing pressure, revealing the drop of carbon coordination from 4 to 3. The broad peaks \textit{d} and \textit{e} at 800-1100~cm\textsuperscript{-1} encode contributions from 3-fold coordinated C atoms (C3). The intensity evolution of this peak with pressure is inverse to that of peak \textit{c}, which further supports the change of coordination of C from 3-fold, at low pressures, to 4-fold at higher pressures. Peak \textit{f} is traced back to C3 sites and, particularly, to C=C double bonds and the reversible intensity drop of this peak upon increasing pressure signals once more reversible C3-to-C4 changes in local structure. The fine structure of the peak \textit{b} around 500~cm\textsuperscript{-1} provides another fingerprint of the population balance between C3 and C4 carbon atoms. In the Raman spectra, right after the transition at 11.2 and 14.5~GPa the peak has a shoulder at around 550~cm\textsuperscript{-1} while at the same pressures peaks \textit{d} and \textit{e} are relatively strong. Upon increasing pressure, the latter two peaks and the shoulder of peak \textit{b} drop and eventually disappear. In figure~S2 (Supp. Mat.), we show the contributions from sulfur bridging two carbon atoms which can be either C3 or C4. The C3-S2-C3 and C4-S2-C3 configurations produce a peak at 550~cm\textsuperscript{-1} while the C4-S2-C4 configuration produces a peak at 500~cm\textsuperscript{-1}. The change of shape of the experimental Raman peak \textit{b} upon compression thus also directly reflects the change in the proportion of C3 and C4 carbon atoms.

\begin{figure*}
\centering
  \includegraphics[height=8cm]{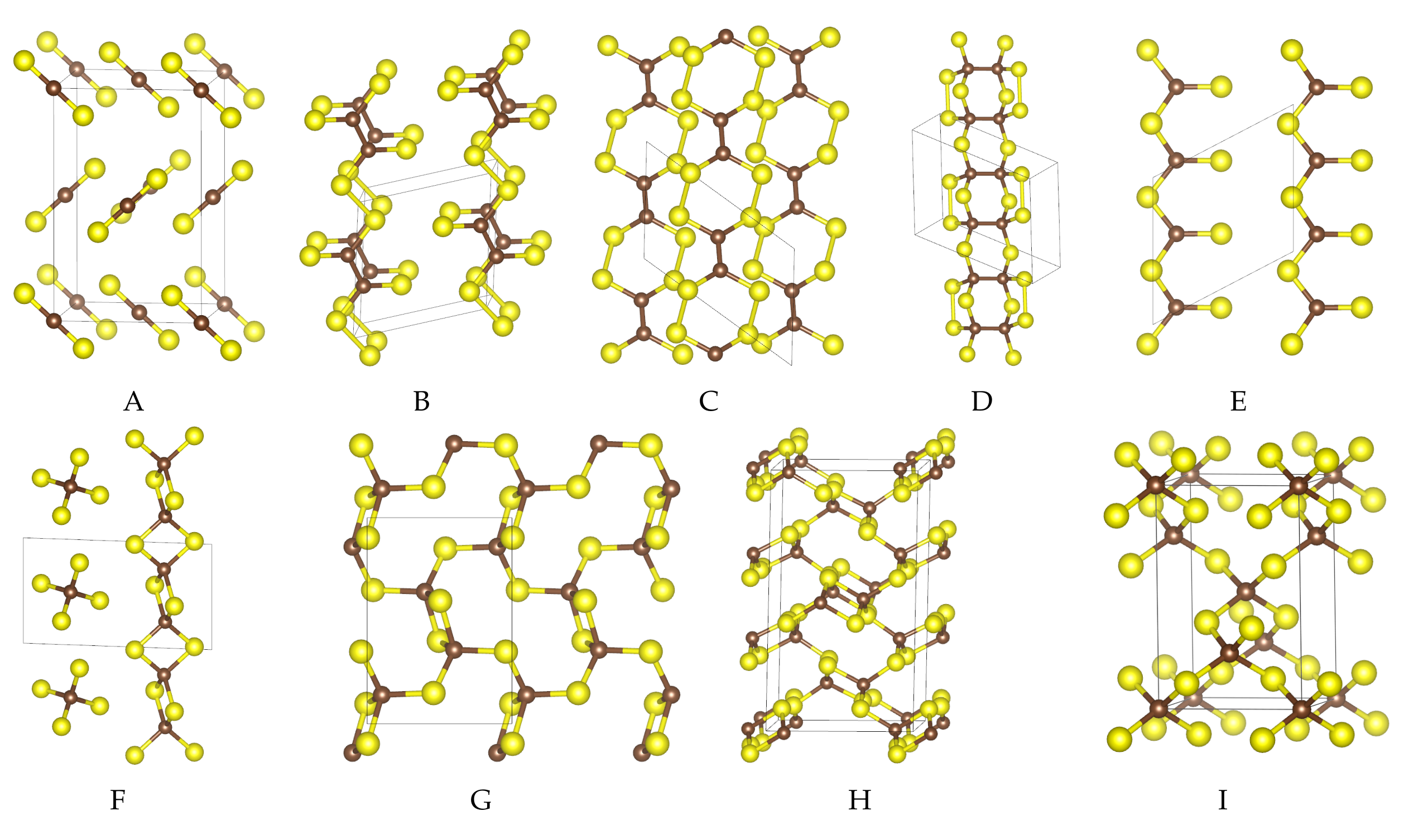}
  \caption{Selected structures found by evolutionary search. A: \textit{Cmce} molecular crystal, B: Frapper polymer, C: double Frapper polymer, D: 262 chain, E: Bridgman’s black polymer, F: \ce{SiS2} NP like chain, G: shahabite, H: \textit{P2\textsubscript{1}/c}, I: $\beta$-cristobalite. A is molecular, B through F are linear polymers, G is layered, H and I are extended systems. In B, C, and E all carbon atoms are in threefold coordination, whereas in D, F, G, H, and I they are all in fourfold coordination.}
  \label{fig:4}
\end{figure*}

\begin{figure}
\centering
  \includegraphics[width=0.95\linewidth]{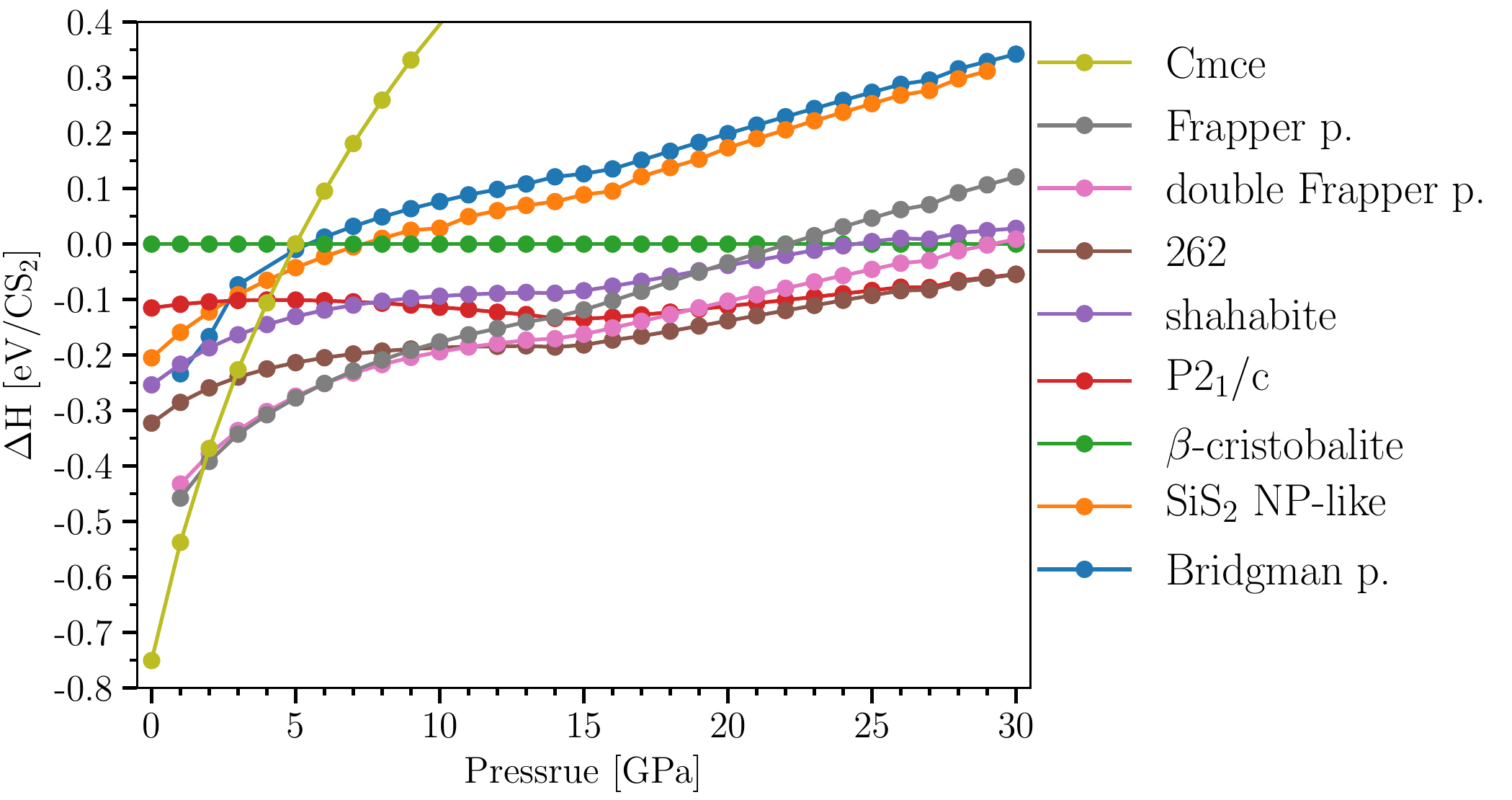}
  \caption{Enthalpy vs. pressure for selected crystal structures of polymeric \ce{CS2} found with evolutionary search. Enthalpy is relative to the $\beta$-cristobalite structure.}
  \label{fig:5}
\end{figure}

For providing a clue to the origin of the pressure evolution in the structure of disordered polymeric \ce{CS2}, we decided to performed evolutionary search of crystal structures of stoichiometric \ce{CS2} in the low-pressure range of 5-10~GPa, not covered in previous studies \cite{Zarifi, Naghavi2015}. The benefit expected from this search is twofold. First, amorphous structures are typically made of a mixture of basic structural blocks of stable and metastable crystalline structures at given pressure. Finding the low-enthalpy crystal structures therefore allows to identify spatial arrangements consisting of ideal structural features. Second, even more valuable understanding is provided by thermodynamics, which allows to assess which features can be expected to be prevalent in equilibrium at a given pressure, thus rationalizing the pressure evolution trends. In the former work by Naghavi et al. \cite{Naghavi2015}, a constrained search was performed targeting the tetrahedral structures in the 0-170~GPa pressure range, identifying fully tetrahedral shahabite \textit{P21/c} as being the most stable structure (snapshot shown in figure~\ref{fig:4}G). In ref.~\cite{Zarifi} instead, fully unconstrained search was performed at 2, 60 and 100~GPa allowing for partial decomposition of the sample and spatial separation of the elements.  Here we focus specifically on the lower pressures around 10~GPa where polymerization occurs in order to understand to which extent the experimental data could be explained by polymerization, without assuming disproportionation. The resulting crystal structures of the search are shown in figure~\ref{fig:4} and their enthalpies vs. pressure are reported in figure~\ref{fig:5}. We first note that already at 5~GPa the enthalpies of \textit{Cmce} molecular crystal (fig.~\ref{fig:4}A), C3 Bridgman polymer BP (fig.~\ref{fig:4}E) and tetrahedral C4 $\beta$-cristobalite structure (fig.~\ref{fig:4}I) cross each other. Surprisingly, the widely accepted BP is not a good structure at any pressure and beyond 5~GPa it becomes the second worst structure right after the \textit{Cmce} molecular crystal while all other structures containing C4 atoms have lower enthalpy. This explains the presence of C4 atoms right after polymerization at around 11~GPa and challenges the traditional view of non-molecular \ce{CS2} being a Bridgman polymer where all C atoms are in 3-fold coordination. We also note that various \ce{CS2} oligomers were studied in ref.~\cite{Frapper2000} where it was concluded that a hypothetical condensed \ce{CS2} phase may present four-connected carbon atoms (oligomer 12b). The lowest enthalpy structure in the low pressure range of 2–5~GPa is the polymer 11c predicted by Frapper, here named Frapper polymer (snapshot shown in fig.~\ref{fig:4}B) \cite{Frapper2000} with volume drop of 21\% from the \textit{Cmce} molecular crystal. A closely related structure arises by bonding remaining terminal sulfurs together, thus creating chain with alternating C=C bonds and double S-S bridges, which we call here double Frapper polymer (see fig.~\ref{fig:4}C). This polymer is thermodynamically preferred in the 6-11~GPa pressure range, although the difference between the single and the double Frapper polymers is very small, i. e. less than 0.01~eV/molecule. Carbon bond length is 1.46~\r{A} and 1.37~\r{A} in former and latter polymer, respectively. The high stability of these polymers featuring direct bonds between two atoms of the same element rather than C-S bonds clearly reflects the metastability of the \ce{CS2} molecule and its tendency to decompose. At the same time, the double Frapper polymer allows us to explain the origin of peak \textit{f} in Raman and IR spectra. The presence of C=C bond results in the VDOS peak at 1400~cm\textsuperscript{-1} (see fig.~\ref{fig:1}) suggesting that the peak \textit{f} can be explained as resulting from plain polymerization (albeit with a different orientation of molecules), with no need for chemical disproportionation. The fact that this is only rarely observed in our simulations is likely an artefact of the short MD simulation times. It is also natural that peaks \textit{e} and \textit{f}, which correspond to C=S and C=C double bonds, disappear upon increasing pressure. On the other hand, the fact that peak \textit{f} reappears even more strongly upon decompression reflects the thermodynamic instability towards decomposition, which may indeed partially occur upon decreasing pressure. The VDOS calculated for the crystal structures are shown in fig.~S3 (Supp. Mat.) and provide further support for the assignment of the Raman and IR peaks, in agreement with our conclusions based on the disordered sample.

The presence of both C3 and C4 carbon atoms right after polymerization at about 11~GPa can be further rationalized by looking for distinct potential polymerization pathways by inspecting close intermolecular C$\cdot\cdot\cdot$S distances in the starting \textit{Cmce} molecular crystal (fig.~S4A Supp. Mat.). It can be seen that molecules with closest C$\cdot\cdot\cdot$S distances form layers in (001) planes (left panel) where C atoms have a distance of 3.12~\r{A} (at 11~GPa) to four S atoms from neighboring parallel molecules within the same layer. This arrangement is ready to produce different intra-layer polymeric chains through at least three distinct transformation paths (fig.~S4 \textit{b}, \textit{c}, \textit{d} Supp. Mat.). While other mechanisms creating bonds between layers are possible as well, they appear to be less likely as C$\cdot\cdot\cdot$S distances across layers are slightly larger. Clearly, the formation of extended covalent network rather than simple chains is possible as well. The existence of several concurrently operating competing polymerization mechanisms provides a plausible explanation for the disorder in the polymeric form since the formation of a polymeric crystalline phase would instead require one mechanism being dominant, which appears to be unlikely.

\section*{Conclusion}
Our combined experimental and computational investigation provides a new scenario for the pressure-induced polymerization of \ce{CS2}. The structure of the disordered polymeric system is more complex than previously thought. It consists of a mixture of C3 and C4 atoms forming a disordered polymeric network with an increasing vs. pressure C4/C3 ratio, which partially retains some crystalline order of S atoms from the molecular crystal. At the polymerization pressure threshold, 10-11~GPa, C4 atoms are thermodynamically preferred to C3 atoms and Bridgman polymer-like chains thus may appear for kinetic reasons only. The electronegativity of C and S is very similar opening more possibilities for polymerization which can occur via C-S bonds as well as via some C=C and S-S bonds. The changes induced by compression are only partially reversible upon decompression, in contrast to the recently observed behaviour in \ce{SO2} \cite{SO2-Zhang}. Importantly, structural transformations observed upon increasing pressure up to 40~GPa can be explained without assuming chemical disproportionation. The structural disorder naturally arises from the presence of a number of possible distinct and competing polymerization pathways in the \textit{Cmce} molecular crystal, suggesting that synthesis of a polymeric \ce{CS2} in crystalline form remains a challenge.

\section*{Methods}
\subsection*{Experimental methodology.}
Liquid \ce{CS2} (purity $\geq$ 99.9\% from Aladdin) was loaded into diamond anvil cells (DACs) at room temperature. We performed Raman spectroscopy using a state-of-the-art confocal Raman microscope with 15 and 2~$\mu$m of axial and transverse resolution, respectively. The spectrometer consisted of a Spectra Pro 750-mm monochromator, equipped with a Pixis Princeton Instrument charge-coupled device detector. Bragg grate filters were used to attenuate the laser light and spatial filtering of the collected light to obtain high-quality spectra down to 7~cm\textsuperscript{-1} with minimal background from the diamond anvils and strong signal from the sample. The laser beam was expanded and cleaned by a band-pass filter. We used a Laser Torus at 660~nm and a Laser Ventus at 532~nm from Laser Quantum. We generally used a 300-grooves-per-mm grating, as the spectral features were getting very broad and weak with pressure. The pressure was determined by the fluorescence of a ruby chip \cite{Mao} placed in the sample chamber or from the stressed part of the diamond anvil \citep{Akahama}. IR absorption spectra have been measured at the Hefei Synchrotron Radiation Laboratory (HESYRL), using a Bruker spectrometer (Vertex 80v) equipped with a Hyperion microscope. The XRD measurements were made at SP-ring~8 (proposal ID:~2019A1363) using a monochromatic X-ray beam with $\lambda$=0.41299~\r{A}, and the scattered X-rays were detected by a 2D Image Plate detector with 100×100~$\mu$m\textsuperscript{2} pixel size. The excellent transverse spatial resolution allows us to obtain clean diffraction patterns of the sample without the presence of spurious diffraction lines from the metallic gasket. The empty cell subtraction, which is of fundamental importance to obtain reliable measurements of the diffuse scattering from an amorphous or liquid sample in the DAC, has been obtained by measuring the empty cell signal at the end of the decompression run after having removed the recovered sample from the gasket.

\subsection*{Simulations methodology}
We performed a structural search for crystalline phases of \ce{CS2} employing the USPEX package \cite{USPEX1, USPEX2} at pressures of 5 and 10~GPa with and without constraints on C-C and S-S bonds, with four formula units (12 atoms). Ab initio simulations were performed by DFT as implemented in VASP 5.3 and 5.4 codes \cite{VASP1, VASP2}, employing projector augmented wave pseudopotentials (with four and six valence electrons for C and S, respectively) and PBE parametrization of the generalized gradient approximation exchange-correlation functional \cite{PBE}, with a cutoff of 520 eV and 2$\pi$*0.06~\r{A}\textsuperscript{-1} final Brillouin zone sampling resolution. Final relaxation and calculation of enthalpy included parameter-free Tkatchenko-Scheffler correction for dispersive forces \cite{TSHI}. Compression and decompression simulations were performed by 20-ps variable-cell isothermal-isobaric (NpT) simulations with Langevin thermostat and $\Gamma$-point Brillouin zone sampling. We used a time step of 2~fs and friction coefficients of 4.0 and 2.0~ps\textsuperscript{-1} for atomic and lattice degrees of freedom, respectively, and 10,000 m\textsubscript{u} as barostat fictitious mass. Data for velocity autocorrelation function were generated by running 20-ps constant volume microcanocical (NVE) simulation. Total and partial VDOS were computed in the standard way as Fourier transform of mass-weighted velocity autocorrelation function from MD trajectories at pressures from 0~GPa to 30~GPa. Static structure factors S(Q) were calculated by performing Fourier transform of the RDFs from MD trajectories at several pressures along compression and decompression runs. All data referred to in the manuscript are available in the article and in Supplemental Material.

\begin{acknowledgments}
O.T. and R.M. were supported by the VEGA project No. 1/0640/20 and the Slovak Research and Development Agency under Contract No. APVV-19-0371. Calculations were performed at the Computing Centre of the Slovak Academy of Sciences using the supercomputing infrastructure acquired in ITMS Projects No. 26230120002 and No. 26210120002 (Slovak Infrastructure for High-Performance Computing) supported by the Research and Development Operational Programme funded by the ERDF. This work was supported by Youth Innovation Promotion Association of CAS (No.2021446), National Natural Science Foundation of China (Grant Nos. 11874361, 51672279, 51727806 and 11774354), Innovation Grant of CAS (No. CXJJ-19-B08), Science Challenge Project (No. TZ2016001), CASHIPS Director's Fund (Grant No. YZJJ201705) and CAS President’s International Fellowship Initiative Fund (2019VMA0027).
\end{acknowledgments}

\bibliographystyle{apsrev4-2}
\bibliography{cs2-references.bib}

\onecolumngrid
\clearpage
\begin{center}
	\textbf{\large Supplemental Material: High Pressure Structural Evolution of Disordered Polymeric CS\textsubscript{2}}
\end{center}
\setcounter{equation}{0}
\setcounter{figure}{0}
\setcounter{table}{0}
\setcounter{page}{1}
\makeatletter
\renewcommand{\theequation}{S\arabic{equation}}
\renewcommand{\thefigure}{S\arabic{figure}}
\renewcommand{\bibnumfmt}[1]{S#1}
\renewcommand{\citenumfont}[1]{S#1}

\vspace{3cm}

\section{Ab initio molecular dynamics compression and decompression}
\begin{figure*}[tbhp]
	\centering
	\includegraphics[width=.95\linewidth]{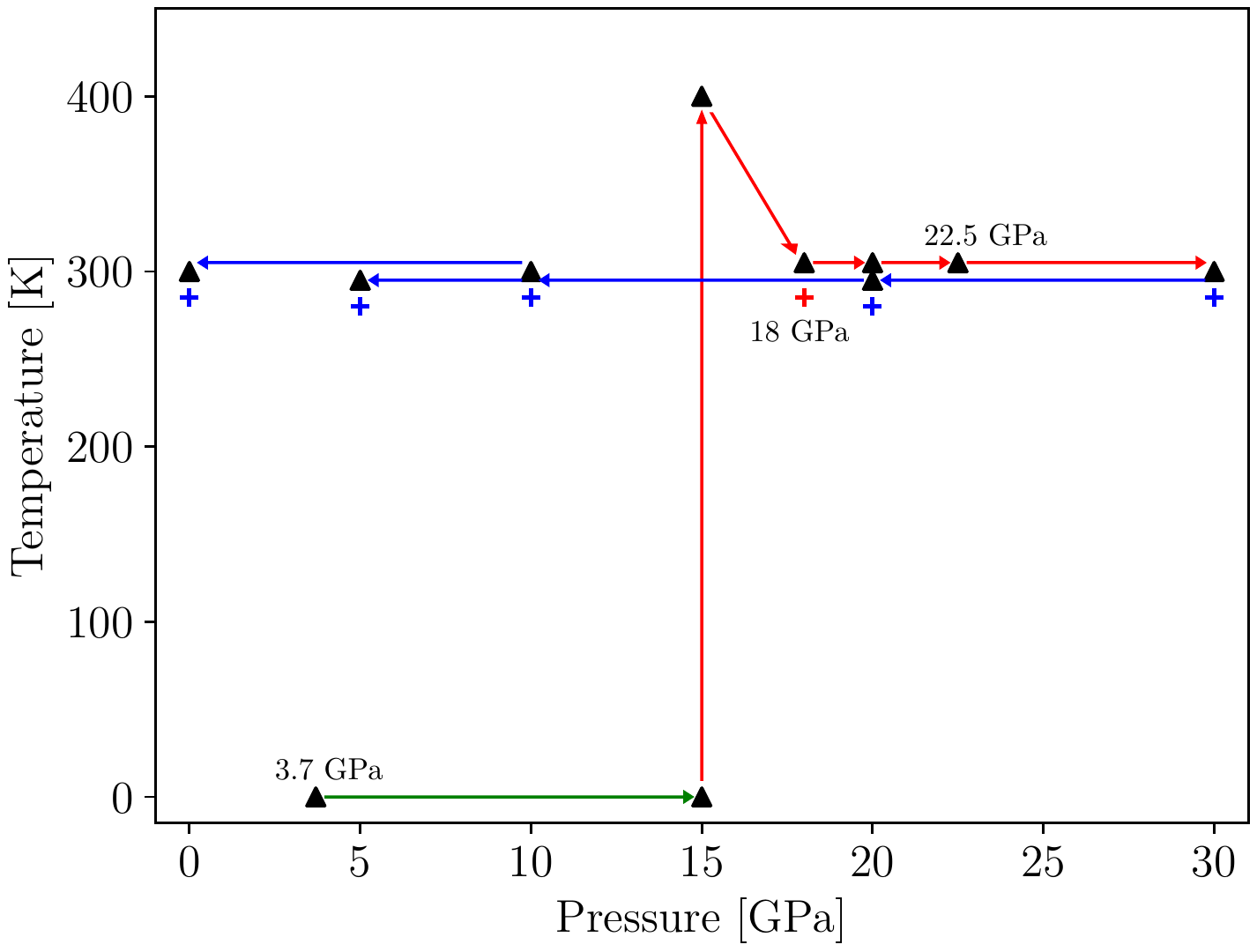}
	\caption{Simulation protocol: initial relaxation of experimental structure to 15~GPa (green arrow), series of compression (red arrows) and decompression (blue arrows) simulations. Plus symbols mark (p,T) points where we performed NVE simulations for calculation of VDOS and $S(Q)$.}
	\label{fig:simulation_protocol}
\end{figure*}

The full simulation protocol of \textit{ab initio} compression and decompression is shown in figure~\ref{fig:simulation_protocol}. We prepared sample from experimental \textit{Cmce} crystal structure at 3.7~GPa by creating $(2 \times 4 \times 2)$ supercell (64 molecules of CS\textsubscript{2}, equivalent to 192-atom unit cell) and by performing structure relaxation to 15~GPa (green curve). Afterwards, compression and decompression simulations were performed by 20-ps variable-cell isothermal-isobaric (NpT) simulations, marked by red and blue arrows, respectively. Sample abruptly polymerized in compression run from 22.5~GPa to 30~GPa, with temperature spiking almost 800~K, indicating it was considerably overpressurized.

\clearpage
\section{Partial VDOS}
\begin{figure*}[tbhp]
	\centering
	\includegraphics[width=.35\linewidth]{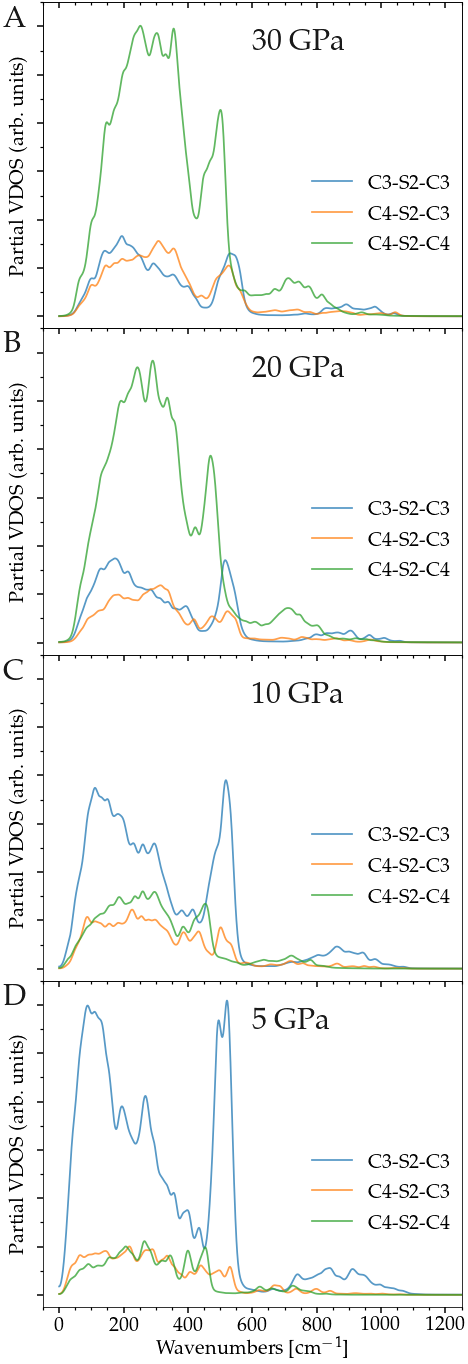}
	\caption{Partial VDOS of sulfur atoms bridging two carbon atoms, calculated at pressures 30, 20, 10 and 5~GPa, labelled A through D, respectively. 3-coordinated carbon atoms are labelled as C3, 4-coordinated as C4. Panels have identical scale. We observe relative exchange in intensity of C4-S2-C4 and C3-S2-C3 peaks, revealing conversion of 4-fold to 3-fold carbon during decompression.}
	\label{fig:sulfur-pvdos}
\end{figure*}

Partial VDOS of bridging sulfur atoms were calculated during decompression of amorphous sample at pressures 30, 20, 10 and 5~GPa and are shown in figure~\ref{fig:sulfur-pvdos}. VDOS of selected crystalline phases found by EA calculated at 20~GPa are shown in figure~\ref{fig:pvdos}. Chemical bonding of selected phases is highlighted by inserts in each figure.

\begin{figure*}
	\centering
	\includegraphics[width=.4\linewidth]{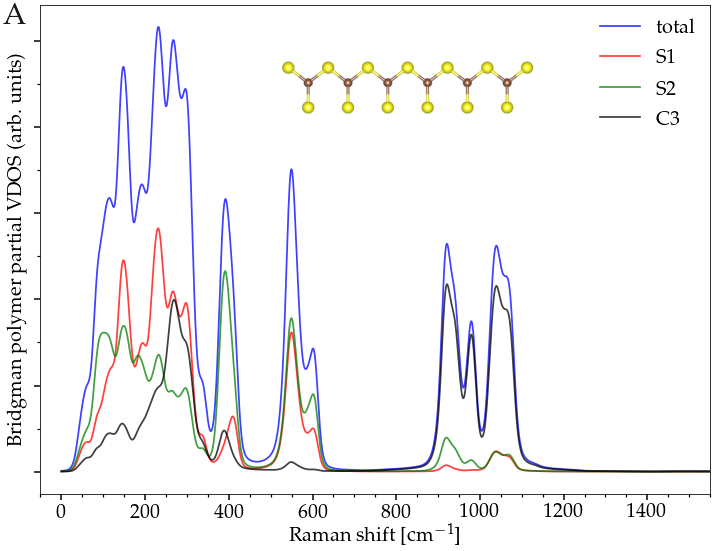}
	\includegraphics[width=.4\linewidth]{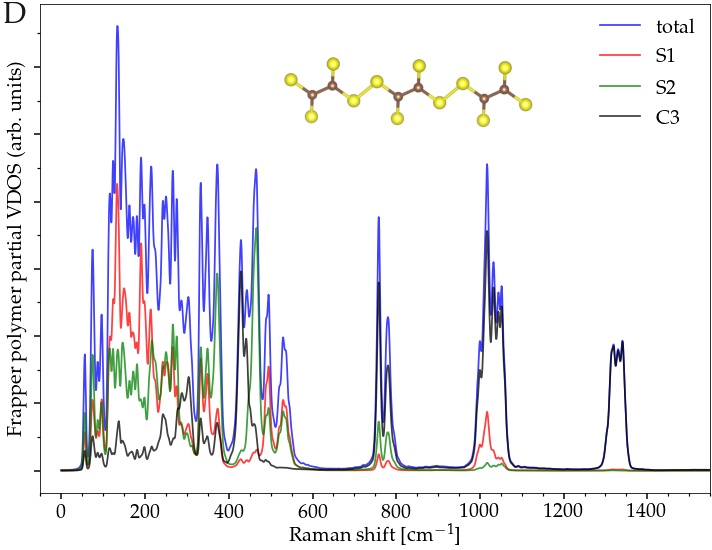}
	\includegraphics[width=.4\linewidth]{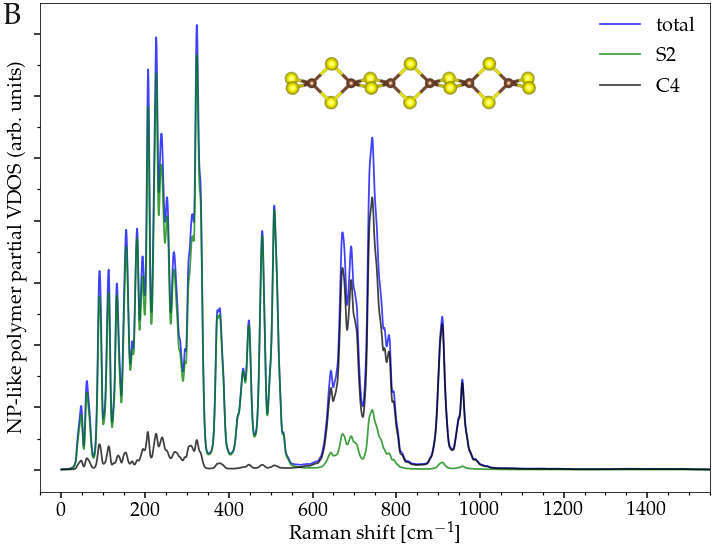}
	\includegraphics[width=.4\linewidth]{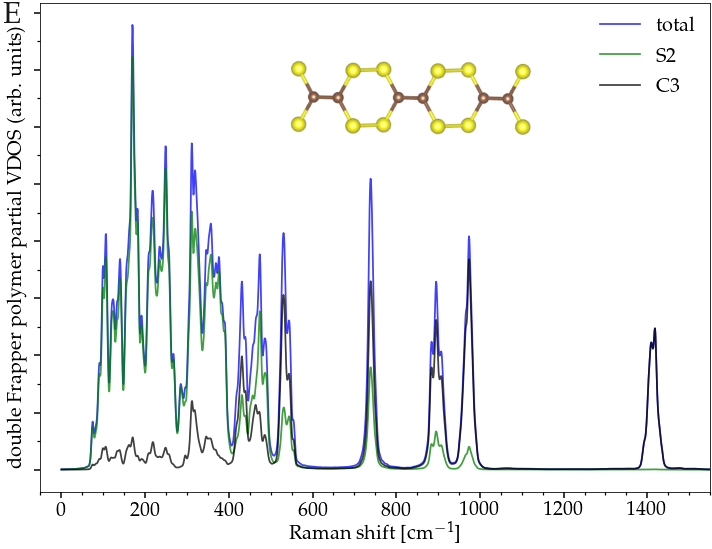}
	\includegraphics[width=.4\linewidth]{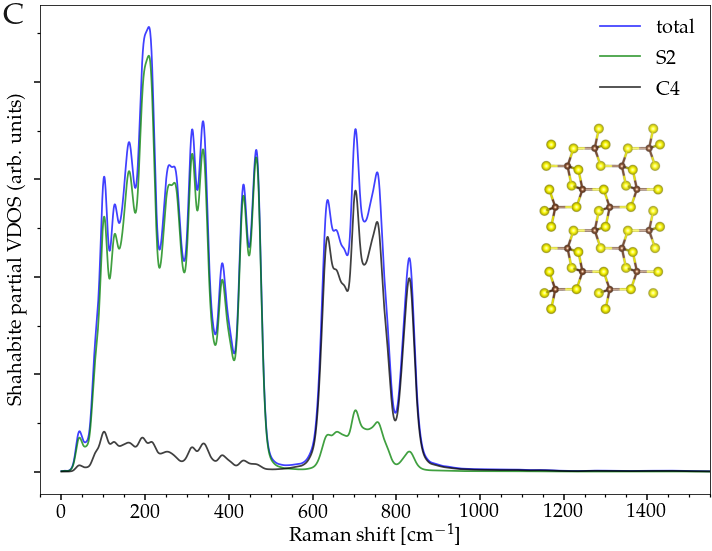}
	\includegraphics[width=.4\linewidth]{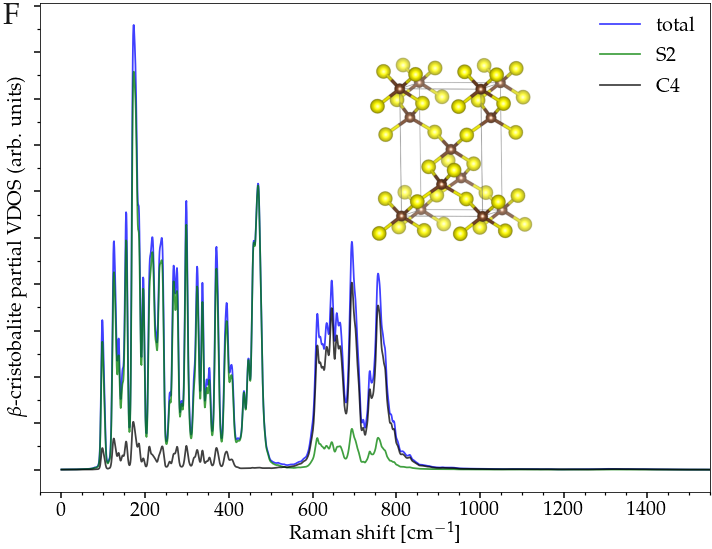}
	\caption{Partial VDOS from atoms with specific structural features; S1: 1-coordinated sulfur (red), S2: 2-coordinated sulfur (green), C3: 3-coordinated carbon and C4: 4-coordinated carbon (both black). Experimental bands \textit{a}, \textit{b}, \textit{c}, \textit{d}, \textit{e} and \textit{f} are at 130~cm\textsuperscript{-1}, 490~cm\textsuperscript{-1}, 740-750~cm\textsuperscript{-1}, 870~cm\textsuperscript{-1}, 1060~cm\textsuperscript{-1} and 1480~cm\textsuperscript{-1}.}
	\label{fig:pvdos}
\end{figure*}

\clearpage
\section{Polymerization mechanism}
Possible pathways for polymerization of \textit{Cmce} molecular crystal could be inferred from inspection of intramolecular distances of neighboring molecules. At polymerization pressure of 11 GPa, each carbon has four sulfur atoms from other molecules at distance of 3.12~\r{A}, implying several possible options for linking CS\textsubscript{2} molecules to chains. The first one is zig-zag pattern in \textit{b} (010) direction, where two molecules connect in alternating (110) and (-110) direction. This linking mechanism may lead to both BP and SiS\textsubscript{2} NP like chains (NP), where NP can be formed either directly or from BP by further polymerization of terminal sulfur atoms, shown in figure~S4B. After linking molecules to BP or NP, carbon atoms can move towards the line between two original molecular columns. Second mechanism follows a diagonal pattern (fig.~S4C), linking molecules to either BP or NP chain in (110) or (-110) direction. This way carbon atoms from molecules are already prepared in line for BP or NP. Third one is again zig-zag pattern, but molecules link in the \textit{a} (100) direction, which leads only to BP (fig.~S4D) while NP is obviously not possible in this case. All mentioned idealized mechanisms operate solely within a layer and therefore would in principle result in a layered structure.

\begin{figure*}[tbhp]
\centering
	\includegraphics[width=.75\linewidth]{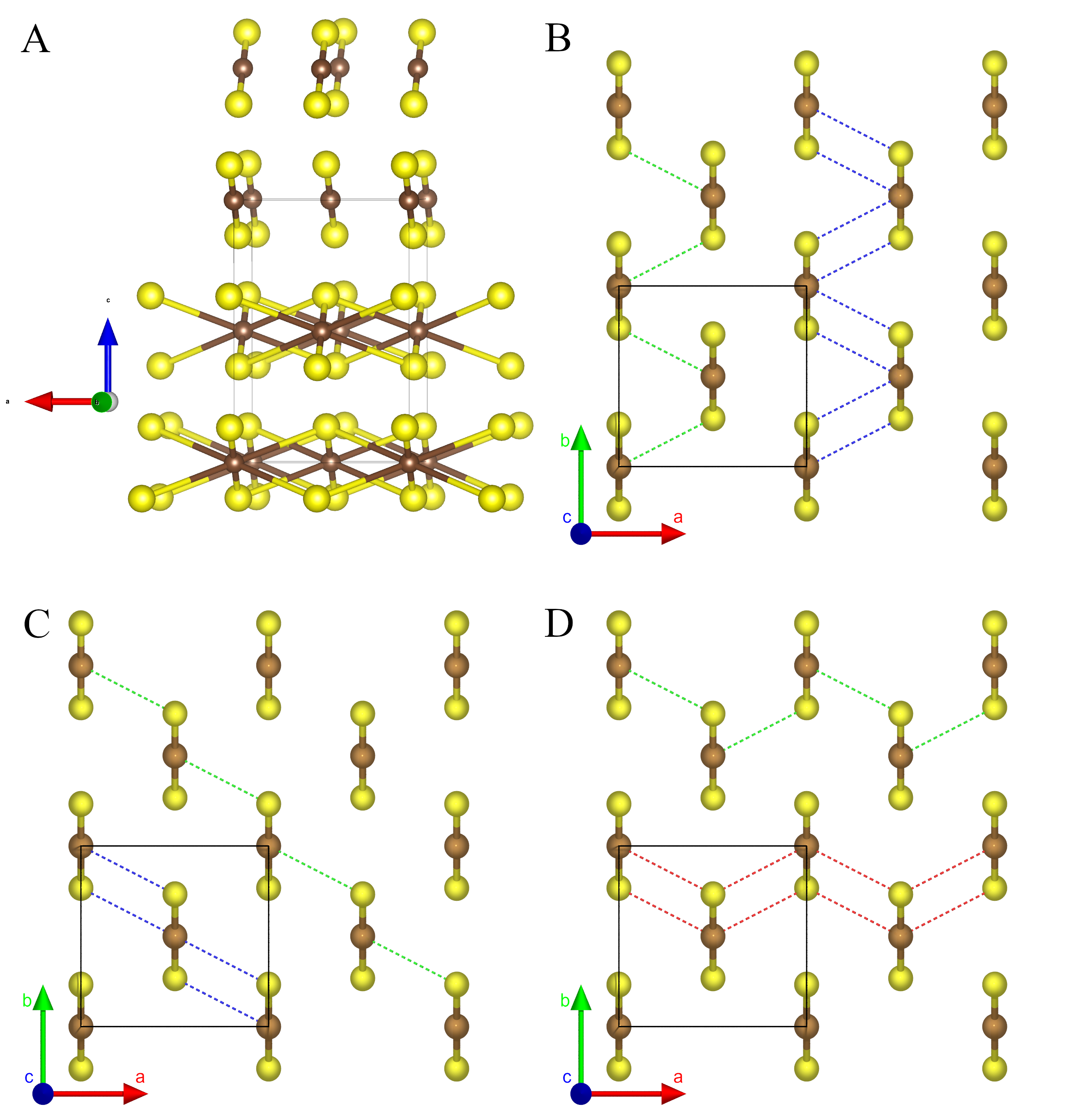}
	\caption{\textit{Cmce} molecular crystal bonding mechanism, based on 11~GPa structure. A: \textit{Cmce} molecular crystal (conventional cell) where in the two bottom layers bond length threshold was artificially increased in order to visualize the closest intermolecular distances. B: zig-zag linking of molecules in \textit{b} (010) direction to BP (green) or NP (blue). BP can easily transform to NP by further polymerization. C: diagonal linking leading to BP (green) or NP (blue). D: zig-zag linking of molecules in \textit{a} (100) direction leading only to BP. Red lines show closest neighbors and highlight why NP chain cannot be formed in \textit{a} direction.}
	\label{fig:pvdos}
\end{figure*}

\FloatBarrier

\end{document}